\documentclass{article}
\usepackage{spconf,amsmath,epsfig}
\usepackage{subcaption}

\title{Cyclic Imaging for All-Sky Interference Forecasting with Array Radio Telescopes}

\name{Gregory Hellbourg and Ian Morrison}
\address{International Centre for Radio Astronomy Research\\
Curtin Institute of Radio Astronomy\\
Perth, Western Australia\\
gregory.hellbourg@curtin.edu.au}

\begin{document}

\maketitle

\begin{abstract}
Radio Frequency Interference (RFI) is threatening modern radio astronomy. A classic approach to mitigate its impact on astronomical data involves discarding the corrupted time and frequency data samples through a process called flagging and blanking. We propose the exploitation of the cyclostationary properties of the RFI signals to reliably detect and predict their locations within an array radio telescope field-of-view, and dynamically schedule the astronomical observations such as to minimize the probability of RFI data corruption.
\end{abstract}

\section{Introduction}

Observational radio astronomy aims at conducting sensitive observations at radio frequencies to detect the faint emissions of astronomical objects using radio telescopes \cite{kraus1986radio}.
The collected data are often corrupted by the presence of strong human emissions referred to as Radio Frequency Interference (RFI) \cite{acevedo1997radio}. The impact of RFI to astronomical data ranges from information contamination (possibly masking the eventual signal of interest) to instrumental deterioration. RFI mitigation is an active field of research, but leads inevitably to a significant loss of information, deteriorating in turn the sensitivity and the productivity of the instrument.

To avoid these repercussions, we propose the development of a real-time all-sky RFI monitor for phased array radio telescopes providing at all times an image of its field-of-view focusing on artificial cyclostationary signals. The collected information on the various sources of RFI enables real-time spectral and spatial environment monitoring, the identification of the most threatening types of RFI, and can open the way to smart dynamic scheduling for minimizing the impact of RFI on the telescope data.

\section{RFI mitigation in radio astronomy}

\subsection{RFI in radio astronomy}

\begin{figure}
\centering
\includegraphics[width=\columnwidth,height=2in]{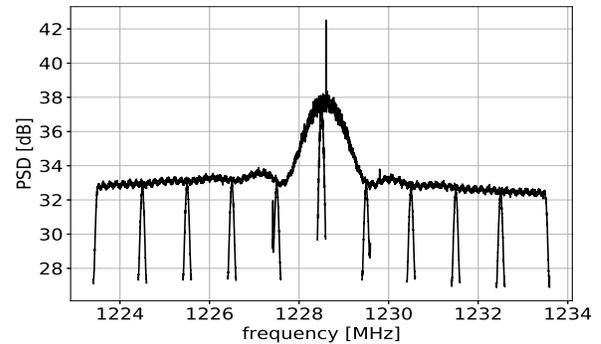}
\caption{GPS satellite signal, as seen by the ASKAP telescope at the Murchison Radio Observatory.}
\label{fig:gps}
\end{figure}

RFI is defined by the International Telecommunication Union as being an ``unwanted but detectable portion of a desired [astronomical] observation that has the potential to either degrade or inhibit the successful conduct of the observation'' \cite{itu}.
An example of RFI corruption is given in Figure \ref{fig:gps}, which shows the transmission of a Global Positioning System (GPS) satellite as received by the ASKAP telescope \cite{deboer2009australian}, located in the Murchison Radio Observatory (Western Australia).


RFI can have various impacts on the successful conduct of an astronomical observation, depending on the telescope receiver sensitivity:

\begin{itemize}
\setlength{\itemsep}{1pt}
\setlength{\parskip}{0pt}
\setlength{\parsep}{0pt}
\item Data masking or corruption preventing the recovery of the astronomical signal of interest.
\item Mislead of the data interpretation due to the possible mimicking of legitimate astronomical signals.
\item Significant loss of data, leading to a reduction of the availability and sensitivity of the instrument (related to the on-sky measurement time), and an increase in operational cost.
\item Inaccuracy in instrumental calibration solutions, for instance when sources of RFI appear as additional point sources in the field of view of the telescope. Time-critical calibration for transient sources is also affected by continuous or even intermittent RFI.
\item Hardware impairment due to telescope receiver saturation.
\end{itemize}

\subsection{RFI mitigation}

\subsubsection{Passive RFI mitigation}

Radio astronomy benefits from international, national, and/or local legal protections to reduce the threat that RFI represents:
\begin{itemize}
\setlength{\itemsep}{1pt}
\setlength{\parskip}{0pt}
\setlength{\parsep}{0pt}
\item Frequency allocation : the usage of the electromagnetic spectrum is regulated at international and local levels to prevent users from overlapping in the frequency domain and interfering with each other. Most of the radio spectrum is however already oversubscribed, and remains unavailable for passive astronomy. Some well-studied spectral segments are protected for radio astronomy, but interference at those frequencies might still occur due to spectral leakage from ``neighbor'' users or harmonic intermodulation products due to non-linearities in astronomical receivers.
\item Radio Quiet Zones (RQZs) : the establishment of RQZs is mostly the responsibility of national entities, and offers legal protections to radio observatories against nearby emitters. Interference from moving emitters can however not be avoided. Ionospheric ducting events also often reflect strong distant emitters unconstrained by the RQZ.
\end{itemize}

\subsubsection{Active RFI mitigation}

Signal processing offers various solutions to mitigate the impact of RFI on astronomical data \cite{baan2011rfi}. Notch filters are often used to reject strong interference occupying narrow frequency bandwidths. Corrupted time-frequency data flagging and blanking is usually a necessary first data pre-processing stage, either achieved automatically based on varying statistical properties of the data, or manually after an expert visual inspection of the data. Spatial filtering methods are emerging as promising solutions for the recovery of uncorrupted time-frequency data when the system architecture permits their implementation \cite{hellbourg2016spatial}.

RFI mitigation never comes at zero cost; it affects the achievable sensitivity of the telescope due to the loss of data, the operational cost when additional equipment is necessary, and the accuracy of the instrumental calibration.
We propose here an alternative solution to prevent these limitations, based on the continuous monitoring and prediction of the dynamic RFI environment of an array radio telescope.

\section{Cyclic all-sky imaging}

\subsection{Data model}

The instantaneous narrowband $M$-elements array radio telescope data model \cite{van2004optimum} is expressed in the following vector form:
\begin{equation}
\mathbf{z}(t) = \mathbf{A_c} \cdot \mathbf{c}(t) + \mathbf{A_r} \cdot \mathbf{r}(t) + \mathbf{n}(t)
\label{vecdatamodel}
\end{equation}
where:
\begin{itemize}
\setlength{\itemsep}{1pt}
\setlength{\parskip}{0pt}
\setlength{\parsep}{0pt}
\item $\mathbf{z}(t) = [z_1(t) \cdots z_M(t)]^T$ is the $M \times 1$ phased antenna array output data vector at time $t$ and  $(.)^T$ is the transpose operator,

\item $\mathbf{c}(t) = [c_1(t) \cdots c_{N_c}(t)]^T$ is the $N_c \times 1$ noise-like stationary astronomical sources signal vector at time $t$,

\item $\mathbf{A_c} = [\mathbf{a}_{c_1}(t, \theta_{c_1}, \phi_{c_1}) \cdots \mathbf{a}_{c_{N_c}}(t, \theta_{c_{N_c}}, \phi_{c_{N_c}})]$ is the $M \times N_c$ astronomical sources space-time signature vectors matrix, with:

\item $\mathbf{a}_{c_n}(t, \theta_{c_n}, \phi_{c_n}) = [a_{c_{n,1}}(t, \theta_{c_n}, \phi_{c_n}) \cdots a_{c_{n,M}}(t, \theta_{c_n}, \phi_{c_n})]^T$ the space-time signature vector corresponding to the $n^{th}$ astronomical source,

\item $\mathbf{r}(t) = [r_1(t) \cdots r_{N_r}(t)]^T$ is the $N_r \times 1$ RFI signal vector at time $t$,

\item $\mathbf{A_r} = [\mathbf{a}_{r_1}(t, \theta_{r_1}, \phi_{r_1}) \cdots \mathbf{a}_{r_{N_r}}(t, \theta_{r_{N_r}}, \phi_{r_{N_r}})]$ is the $M \times N_r$ RFI space-time signature vectors matrix, with:

\item $\mathbf{a}_{r_n}(t, \theta_{r_n}, \phi_{r_n}) = [a_{r_{n,1}}(t, \theta_{r_n}, \phi_{r_n}) \cdots a_{r_{n,M}}(t, \theta_{r_n}, \phi_{r_n})]^T$ the space-time signature vector corresponding to the $n^{th}$ RFI,

\item $\mathbf{n}(t) = [n_1(t) \cdots n_M(t)]^T$ is the $M \times 1$ random independent and identically distributed (i.i.d.) centered system noise vector at time $t$ with stationary complex Gaussian distribution with covariance matrix $\mathbf{R_n}$.

\end{itemize}

\subsection{Cyclostationarity}

Information-bearing telecommunication signals present a hidden periodicity due to the periodic characteristics involved in the signal construction (carrier frequency, baud rate, coding scheme...). These parameters are usually hidden by the randomness of the message to be transmitted. However, by using a cyclostationary approach \cite{gardner2006cyclostationarity} this hidden periodicity can be recovered, thus making the detection of the telecommunication signal possible.

\begin{figure*}[h!]
\centering
\makebox[\textwidth][c]{\includegraphics[width=1.1\textwidth,height=1.4in]{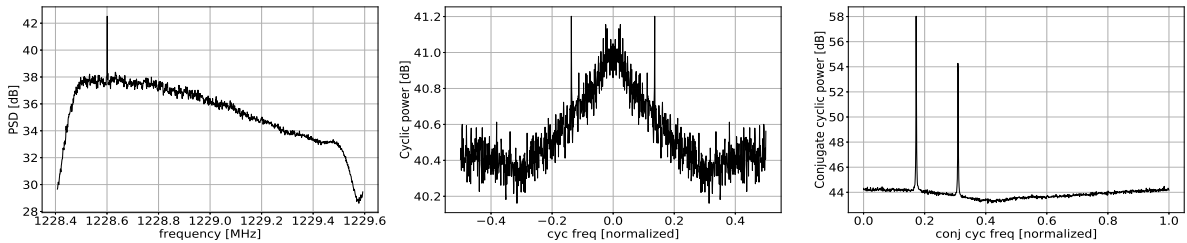}}
\caption{Cyclic analysis of the GPS satellite signal from Figure \ref{fig:gps}. The left plot shows the power spectral density of the satellite signal, and the middle and right plots show the cyclic and conjugate cyclic spectra of the signal, respectively.}
\label{fig:cycliccanalysis}
\end{figure*}

The array correlation matrix
\begin{align*}
\mathbf{R_{z}} &= \left\langle \mathbf{z}(t)\mathbf{z^{H}}(t)\right\rangle _{\infty}\\
&= \mathbf{A_{r}R_{r}A_{r}^H} + \mathbf{A_{s}R_{s}A_{s}^H} + \mathbf{R_n}
\end{align*}
where $(.)^H$ is the Hermitian transpose, $\left\langle.\right\rangle _{\infty}$ is the infinite time average operator, and $\mathbf{R_{s}}$ and $\mathbf{R_{r}}$ are the diagonal matrices containing the individual astronomical and RFI sources power in their main diagonal, respectively -- is replaced by the cyclic correlation matrix: 
\begin{equation}
\mathbf{R_{z}^{\alpha}}=\left\langle \mathbf{z}(t)\mathbf{z^{H}}(t)\exp(-j2\pi\alpha t)\right\rangle _{\infty}\label{eq:cyclic_corr_mat}
\end{equation}
where $\alpha$ is the cyclic frequency. This parameter is related to the above-mentioned periodic characteristics \cite{gardner1987spectral,gardner1987spectral2}. Similarly, we can define the cyclic conjugated correlation matrix, $\overline{\mathbf{R}}_{\mathbf{z}}^{\alpha}$, by replacing the operator $(.)^H$ with the standard transpose $(.)^T$ in equation \ref{eq:cyclic_corr_mat}. Then, another set of cyclic frequencies can be considered. An example of cyclostationarity analysis of the GPS satellite signal shown in Figure \ref{fig:gps} is depicted in Figure \ref{fig:cycliccanalysis}, where the two spectra on the right are the cyclic and conjugate cyclic specta of the GPS signal whose power spectral density is seen on the left plot. Two dominant cyclic frequencies can be identified on the conjugate cyclic spectrum.

Any signal that is cyclostationary at cyclic frequency $\alpha_0\neq0$ will generate non-zero cyclic or cyclic conjugated correlation matrices. Conversely, any stationary signal or cyclostationary signal with different cyclic frequencies will generate zero in equation \ref{eq:cyclic_corr_mat}. Thus, the cyclic correlation matrix at $\alpha_{0}$ for the model becomes:
\begin{equation}
\mathbf{R}_{\mathbf{z}}^{\alpha_{0}}=\mathbf{a}_{r}\mathbf{a}_{r}^{H}R_{\mathbf{r}}^{\alpha_{0}}+\underbrace{\mathbf{A_{s}R_{s}}^{\alpha_{0}}\mathbf{A}_s^H}_{\rightarrow0}+\underbrace{\mathbf{R_n}^{\alpha_{0}}}_{\rightarrow0}\label{eq:CycCorrInterim}
\end{equation}
where $\alpha_{0}$ is the expected (or detected) RFI cyclic frequency and $R_{\mathbf{r}}^{\alpha_{0}}$ is the cyclic power of the cyclostationary RFI. To simplify the notations, RFIs with different cyclic frequencies are considered here as astronomical source signals and are merged into $\mathbf{R_s^{\alpha_0}}$. A similar asymptotic expression can be obtained for the cyclic conjugated correlation matrix.

\subsection{Cyclic imaging}

Phased array radio telescopes, when appropriately calibrated, allow the imaging of their field-of-view through beamforming, i.e. phasing each antenna element of the array to compensate for the geometrical delay between the elements given a particular pointing direction. The field-of-view can then be imaged by measuring the power received in each possible direction within the array's field-of-view, using the array correlation matrix $\mathbf{R_{z}}$, the array geometrical configuration (i.e. relative location of each antenna), and the observation frequency $f_0$:
\begin{equation}
M(x,y) = \mathbf{a}_{f_0,x,y}^H\mathbf{R_{z}}\mathbf{a}_{f_0,x,y}
\end{equation}
where $M(x,y)$ is the skymap pixel matrix, $(x,y)$ are the coordinates of a given direction in the telescope field-of-view as represented by a planar projection of the celestial sphere above the array, and $\mathbf{a}_{f_0,x,y}$ is the phase compensation vector corresponding to the direction $(x,y)$.

\begin{figure}[h!]
\centering
\begin{subfigure}[b]{0.49\columnwidth}
\includegraphics[width=1.1\textwidth]{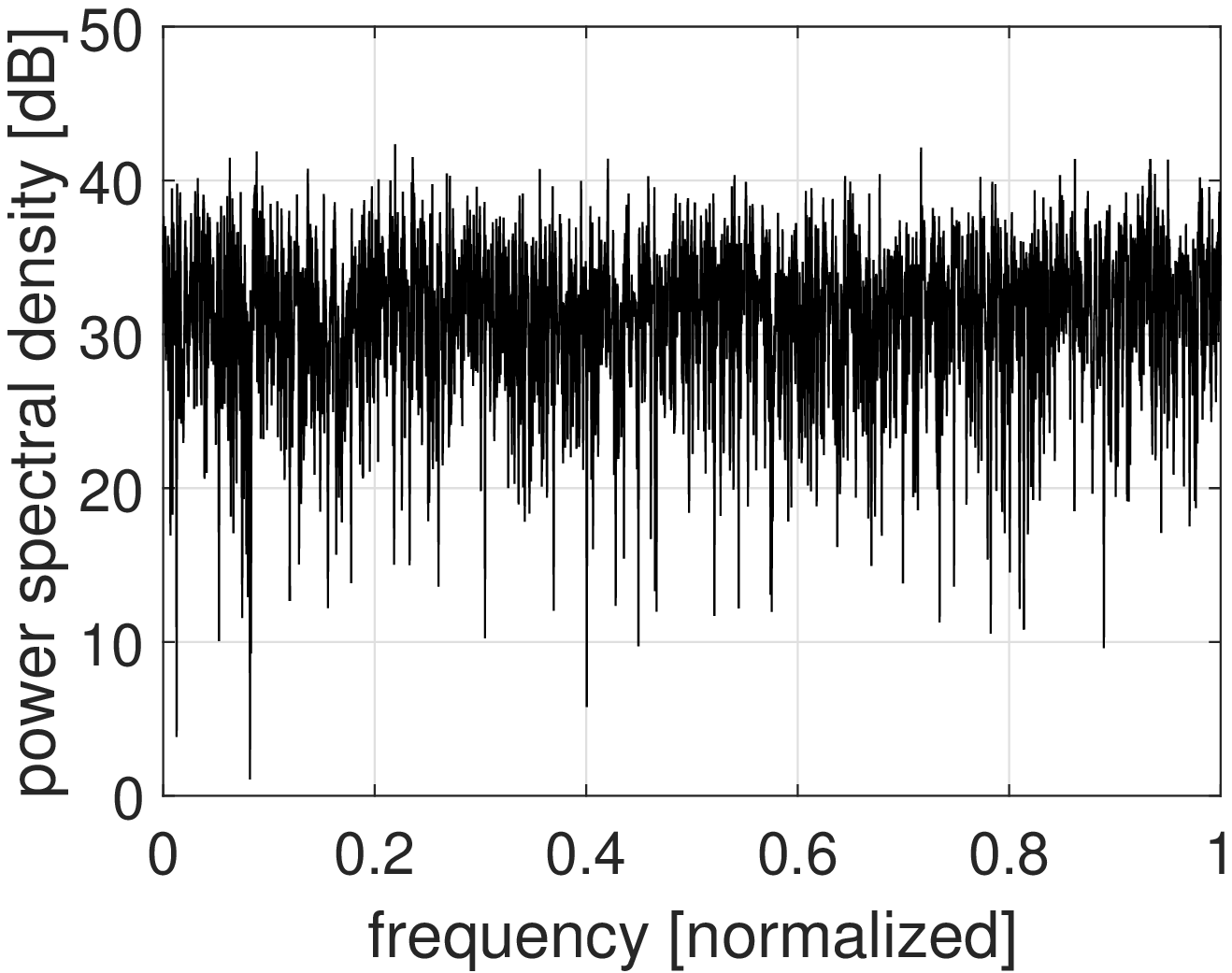}
\caption{}
\label{fig:a}
\end{subfigure}
\begin{subfigure}[b]{0.49\columnwidth}
\includegraphics[width=1.1\textwidth]{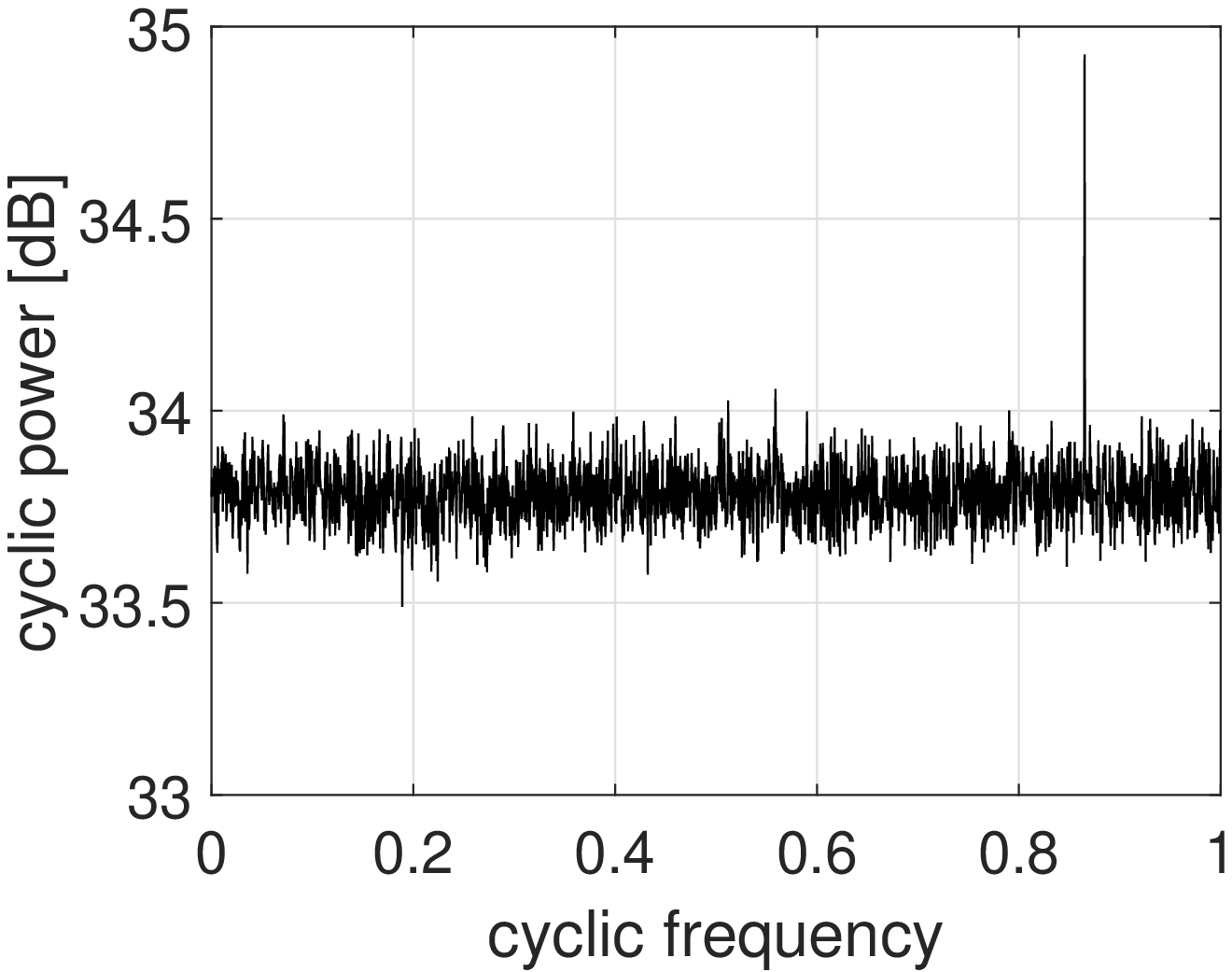}
\caption{}
\label{fig:b}
\end{subfigure}

\begin{subfigure}[b]{0.49\columnwidth}
\includegraphics[width=1.1\textwidth]{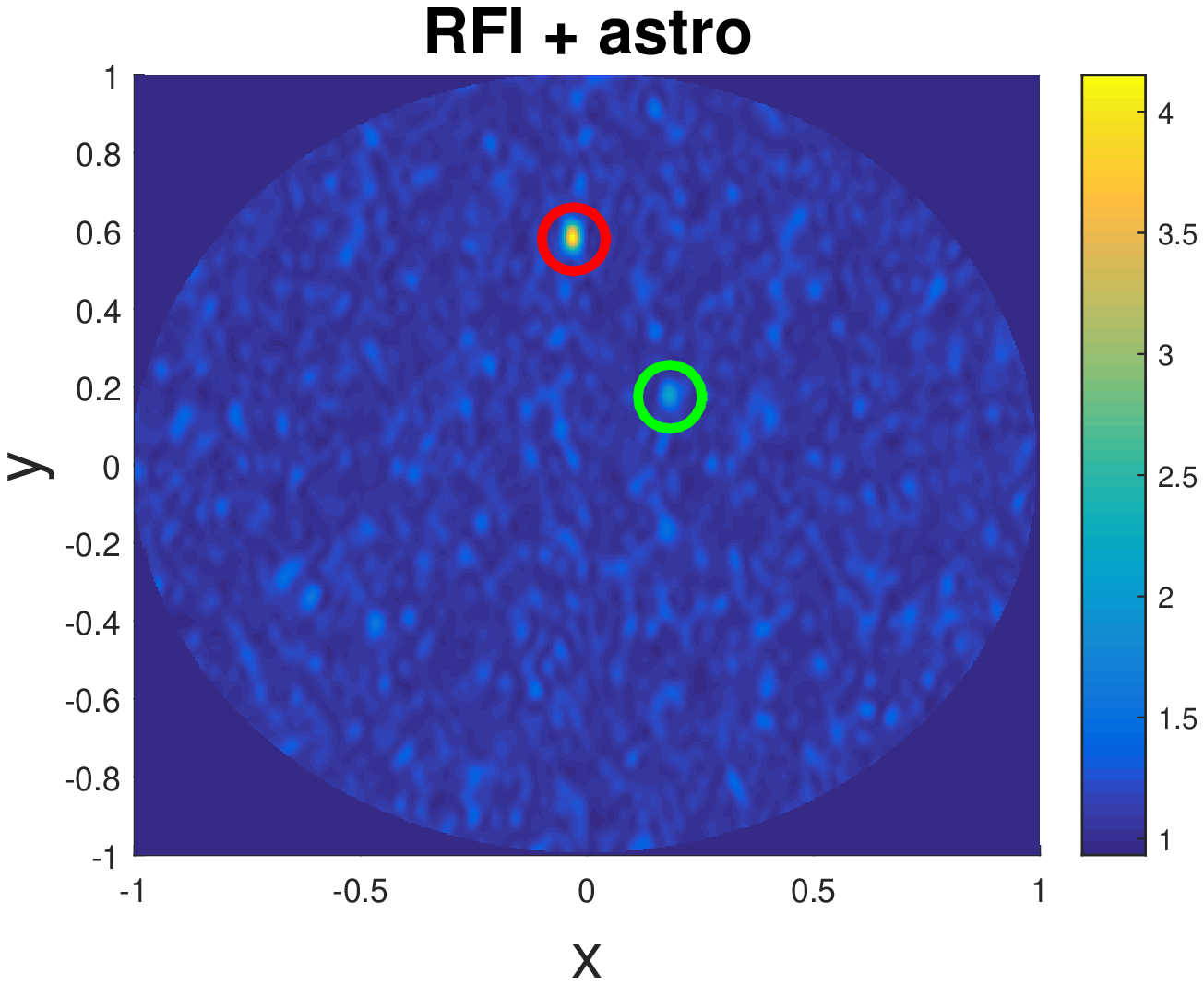}
\caption{}
\label{fig:c}
\end{subfigure}
\begin{subfigure}[b]{0.49\columnwidth}
\includegraphics[width=1.1\textwidth]{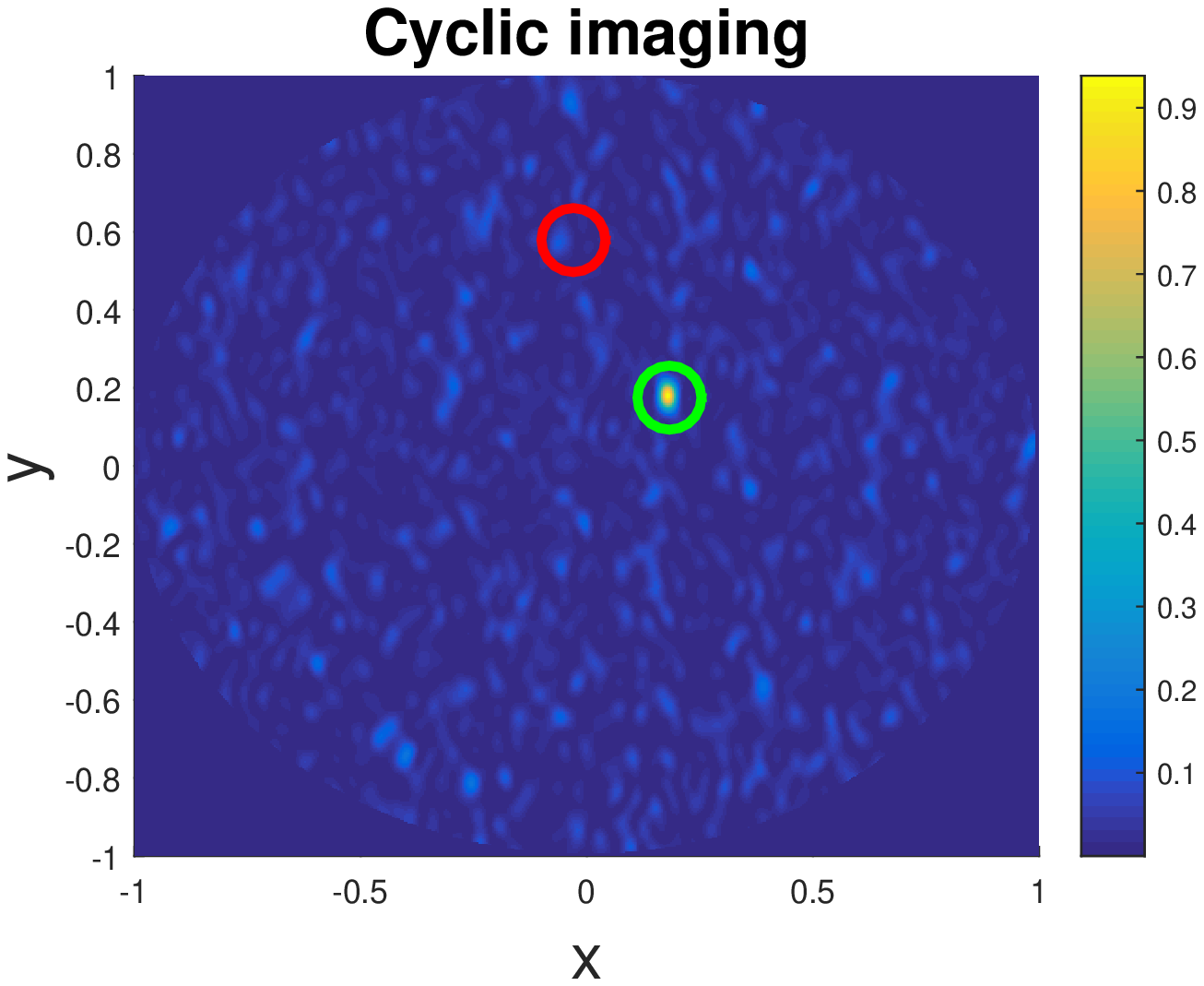}
\caption{}
\label{fig:d}
\end{subfigure}
\caption{Example of cyclostationary imaging with a simulated 48 phased antenna array, 2048 samples. (a) Power spectral density of signal received by one antenna of the array, containing one BPSK signal (SNR=0dB), one astronomical source (SNR=+5dB), and white Gaussian system noise. (b) Cyclic spectrum highlighting one cyclic frequency corresponding to the BPSK signal. (c) Skymap produced with the classical array correlation matrix. The red circle shows the location of the astronomical source, the green circle shows the location of the BPSK emitter. (d) Skymap produced with the cyclic covariance matrix at the cyclic frequency of the BPSK signal. The stationary astronomical source has disappeared.}
\label{fig:cyclo_imaging}
\end{figure}

Figure \ref{fig:cyclo_imaging}.(a) shows the power spectral density of a simulated signal received by a virtual 48-antenna array, consisting of a BPSK signal (SNR = 0dB), a stationary noise-like astronomical source (SNR = +5dB), and stationary system noise. After producing the array correlation matrix of the data, a skymap is produced and can be seen in Figure \ref{fig:cyclo_imaging}.(c). The green and red circles represent the locations of the BPSK and astronomical sources, respectively.

Cyclic imaging follows the same process, but applied to the cyclic covariance matrix as opposed to the classical array covariance matrix \cite{hellbourg2014radio}. Figure \ref{fig:cyclo_imaging}.(b) shows the cyclic spectrum of the data, and a peak associated with the cyclic frequency of the BPSK signal can clearly be identified. A skymap based on the cyclic covariance matrix at this cyclic frequency is shown in figure \ref{fig:cyclo_imaging}.(d). The astronomical source is no longer seen, as the contribution of all stationary sources is here highly attenuated. The BPSK emitter can now be easily and unambiguously detected.

\section{Dynamic telescope scheduling based on cyclic imaging monitoring}

Telescope observation scheduling is based on many environmental and logistical factors, such as the telescope configuration (radio receiver changes, geometrical re-configuration), day/night cycles, weather and ionospheric perturbations..., but never accounts for the RFI environment.
All-sky RFI monitoring enables the tracking and prediction of the dynamic RFI environment of a telescope, and allows optimum observation scheduling based on RFI avoidance to minimise data corruption and maximise the instrument efficiency.

Cyclic RFI monitoring is particularly sensitive to artificial communication signals, as opposed to traditional astronomical instrumentation. It therefore allows the detection of faint sources of RFI, usually undetected by classic automatic and manual data flaggers. These weak RFI are, however, still sources of artifacts appearing after the long time integrations required for sensitive scientific observations. The proposed system will be able to extract statistics of the RFI environment, and the inference from data fusion of the extracted parameters will allow a real-time assessment of the radio spectral and spatial occupancy, but also the short-term (of the order of seconds) and long-term (of the order of hours) temporal, spectral, and spatial RFI occurrences, predicted through extrapolation methods.


Three types of sources of RFI can be distinguished :
\begin{itemize}
\setlength{\itemsep}{1pt}
\setlength{\parskip}{0pt}
\setlength{\parsep}{0pt}
\item Stationary RFI : Spatially stationary sources of interference are avoided by ensuring the telescope does not steer in their direction of arrival. The avoidance strategy accounts for the celestial sphere rotation, and ensures that astronomical observations are conducted when the stationary RFI is out of sight of the telescope.
\item Slowly moving RFI : Slowly moving transmitters can be predicted with high accuracy (of the order of the telescope angular resolution), and the avoidance strategy involves switching to a more appropriate telescope steering direction (i.e. different astronomical project in general) when the current observation presents a high risk of RFI corruption.
\item Rapidly moving RFI : Rapidly moving sources of RFI are complicated to predict, and the instrumental reconfiguration required to avoid these sources of interference is usually time consuming compared to the temporal impact of them on the collected data. The data corruption is often negligible compared to the observation duration (a few seconds of data corruption compared to hours of observation). The proposed system will in this case track the times and frequencies at which RFI corruption happens, depending on the angular separation between the source of RFI and the telescope steering direction, and automatically flag and discard the impacted data. This approach will ensure the reliability of automatic data flaggers in the low signal-to-noise ratio regime.
\end{itemize}

All-sky cyclic imaging provides a real-time imaging and monitoring of all artificial signals impinging on the telescope, with the ability to detect and identify the various sources through their cyclostationary features (cyclic frequencies) associated with time-frequency-space information extracted from the imaging system (location, motion, trajectory/ephemeris, central frequency, time occurrence...)\cite{xu1992direction,xin1998directions}

\section{Conclusion}

We presented here the concept of telescope dynamic scheduling that can account for the dynamic RFI environment to minimize the RFI corruption of astronomical data. The RFI environment is monitored through all-sky cyclic imaging, separating information-bearing signals from stationary astronomical and system noise sources, and improving therefore the accuracy and sensitivity of RFI detection. The parameters extracted from the monitoring system, such as spectral occupancy, cyclic frequencies, location, or motion, are then exploited to predict the RFI environment over short and long time scales. Accounting for the astronomical programs scheduled for the telescope (defined by the observed field in the sky and the frequency span), the monitor will eventually generate a dynamic observing schedule minimizing the RFI corruption of the telescope data. The monitoring data can also be exploited to automatically generate preventative time and frequency flagging masks.


\end{document}